# Modeling genome evolution with a diffusion approximation of a birth-and-death process

Georgy P. Karev[1], Faina S. Berezovskaya[2], and Eugene V. Koonin[1,*]

[1]National Center for Biotechnology Information, National Library of Medicine, National Institutes of Health, Bethesda, MD 20894, and [2]Department of Mathematics, Howard University, 2400 Sixth Str., Washington D.C., 20059, USA.

*To whom correspondence should be addressed. Email koonin@ncbi.nlm.nih.gov

**Abstract**

**Motivation**: In our previous studies, we developed discrete-space Birth, Death and Innovation Models (BDIM) of genome evolution. These models explain the origin of the characteristic Pareto distribution of paralogous gene family sizes in genomes, and model parameters that provide for the evolution of these distributions within a realistic timeframe have been identified. However, extracting temporal dynamics of genome evolution from discrete-space BDIM was not technically feasible. We were interested in obtaining dynamic portraits of the genome evolution process by developing a diffusion approximation of BDIM.

**Results**: The diffusion version of BDIM belongs to a class of continuous-state models whose dynamics is described by the Fokker-Plank equation and the stationary solution could be any specified Pareto function. The diffusion models have time-dependent solutions of a special kind, namely, the generalized self-similar solutions, which describe the transition from one stationary distribution of the system to another; this provides for the possibility of examining the temporal dynamics of genome evolution. Analysis of the generalized self-similar solutions of the diffusion BDIM reveals a biphasic curve of genome growth in which the initial, relatively short, self-accelerating phase is followed by a prolonged phase of slow deceleration. This evolutionary dynamics was observed both when genome growth started from zero and proceeded via innovation (a potential model of primordial evolution) and when evolution proceeded from one stationary state to another. In biological terms, this regime of evolution can be tentatively interpreted as a punctuated-equilibrium-like phenomenon such that whereby evolutionary transitions are accompanied by rapid gene amplification and innovation, followed by slow relaxation to a new stationary state.

# 1. INTRODUCTION

Analysis of the numerous available complete genome sequences yields various quantitative characteristics of genomes, such as the distribution of the number of genes or pseudogenes in paralogous families, the number of transcripts per gene, the number of interactions per protein, the number of genes in coexpressed gene sets, the number of connections per node in metabolic and regulatory networks, and others (1-5). Notably, it has been shown that the distributions of many of these variables are well approximated by a power law or, more precisely, a generalized Pareto function (3,4,6-9). These distributions can be used as inputs for mathematical models that have the potential of revealing non-trivial "laws" of genome evolution.

In several previous studies, we and others attempted to model the evolution of the gene repertoire of genomes within the framework of a birth-and-death process (8,10-13). The set of genes in a genome always can be represented as a collection of paralogous gene families, including families with a single member. A birth-and-death process appears to be a natural formalism for describing gene family evolution. Indeed, by definition, evolution of paralogous families occurs by gene duplication (followed by mutational diversification) which must be the crucial elementary process in any model of genome evolution and can be designated *gene birth*. Obviously, however, genomes do not grow indefinitely, so gene duplication must be counter-balanced by gene elimination; indeed, it has been shown in numerous studies that extensive gene loss is common during evolution. Hence *gene death* is the second elementary process to be included in models of genome evolution. In addition, genes new to a given lineage may emerge as a result of extreme divergence after duplication erasing the memory of a gene's origin, horizontal gene transfer (HGT), and, perhaps, evolution of a protein-coding gene from a non-coding sequence. Collectively, these aspects of genome evolution can be described as *innovation*.

We employed gene birth, gene death and innovation as elementary processes to develop Birth-Death-Innovation models (BDIMs) of genome evolution. Analysis of the deterministic version of BDIMs showed that the characteristic power law asymptotic of the size distribution of gene families is observed if, and only if, birth and death rates of genes in families of sufficiently large size are balanced, i.e., asymptotically equal up to the second order. Furthermore, it has been found that the simplest model compatible with the empirical gene family size distributions is the *linear* $2^{nd}$ order balanced BDIM (10,14). The subsequent development of the stochastic version of BDIMs allowed us to examine not only for the stationary state of the genome but also some of its evolutionary characteristics, including the mean times of formation and extinction of a family of a given size (12,13). Substituting the published estimates of the rates of gene duplication and loss (15) into different versions of the model, we found that the linear BDIM, which gives good approximations of the stationary distributions of family sizes for different species, predicts unrealistically long mean times for the formation of the largest families. As shown by computer simulations, even the minimum time required for the formation of the largest family under the linear BDIM exceeded the time actually available for evolution by orders of magnitude. Thus, the linear BDIM is incompatible with the estimates of the rate of genome size growth derived from the empirical data. Therefore we explored non-linear, higher degree BDIM and showed that the mean time of formation of a gene family under a fixed average duplication rate went through a minimum at model degrees between 2 and 3. Even these mean times were much longer than the allotted time of evolution. However, using Monte Carlo simulations, we showed that the minimum formation time of families of the expected size under BDIMs of the orders between 2-3 fit the timescale of genome evolution (13). The finding that only higher degree BDIMs can adequately describe the evolution of the gene family size distribution indicates that the growth of gene families (at least large ones) is self-accelerating, which might reflect positive selection driving the fixation of gene duplications in such families.

Here, we develop and explore a distinct class of BDIMs, namely, diffusion models whose dynamics is described by the Fokker-Plank equation; the exact stationary solutions of these models are the same Pareto functions (asymptotically tending to a power law) that have been obtained as approximation of the stationary solutions for deterministic and stochastic BDIMs (12-14). We examine the temporal dynamics of the diffusion model and find not only the stationary but also the *time-dependent* solutions. A class of *self-similar,* time-dependent solutions is identified in an exact analytical form; these solutions present dynamic portraits of the formation of the stationary distribution.

The proofs of the main mathematical assertions and theorems involved in this work are appearing elsewhere (16).

## 2. The diffusion approach and the master model

Let a population be subdivided into *N* (finite or infinite) sets which we will call "families" and *f(x,t)* be the number of families of size *x* at moment *t*. Let us suppose that the "individual" birth and death rates in a family of size *x* are $\lambda(x)$ and $\delta(x)$, respectively. Then the equation

$$\frac{\partial f(t,x)}{\partial t} = f(t,x-1)\lambda(x-1) - (\lambda(x)+\delta(x))f(t,x) + \delta(x+1))f(t,x+1)], \quad x=0,\ldots N \quad (2.1)$$

(subject to boundary conditions) describes the birth-and-death process with the set of states {0,1,…*N*}. In particular, if

$$\frac{\partial f(t,0)}{\partial t} = -f(t,0)\lambda(0) + \delta(1))f(t,1), \quad (2.2)$$

$$\frac{\partial f(t,N)}{\partial t} = f(t,N-1)\lambda(N-1) - \delta(N))f(t,N)$$

then we get the birth-and-death process with reflecting boundaries at 0 and *N*. BDIM is a special case of this process, the stochastic behavior of which has been explored in detail in our previous work (13).

The stationary solution of model (2.1), (2.2) is known; this single equilibrium is asymptotically stable and follows asymptotically the power law distribution, $f_{st}(x) \sim x^{-\gamma}$ if the model is 2$^{nd}$ order balanced, i.e., if, by definition, $\lambda(x-1)/\delta(x) = 1-\gamma/x + O(1/x^2)$ for large *x* (10).

The temporal dynamics of birth-and-death model is known in explicit analytical form only for some specific cases although system (2.1) can be easily solved numerically; the only technical problem is the high dimensionality of the system. It is well known that birth-and-death process with large number of states can be approximated under certain conditions by a diffusion process with a continuous space of states, and vice versa (e.g., (17,18)). In many cases, the analytical and qualitative analysis of the diffusion model is easier than the analysis of the original, discrete birth-and death process. Let us consider the continuous-space approximation of model (2.1) (with finite and infinite *N*). Instead of the discrete space of states {0,1,2,...N}, we consider a continuous space of states, the interval [r,N]. A choice of the non-negative minimal possible value of x, r, (e.g., r=0 or r=1), depends on the statement of a problem. Formally approximating the right-hand side of the system (2.1) by Taylor's expansion and truncating at second derivatives (see, e.g., (19)), we obtain the master equation:

$$\frac{\partial f(t,x)}{\partial t} = -\frac{\partial}{\partial x}[f(t,x)\,\mu(x)] + \frac{1}{2}\frac{\partial^2}{\partial x^2}[f(t,x)\,\sigma^2(x)] \qquad (2.3)$$

where $\mu(x)=\lambda(x)-\delta(x)$ is the drift coefficient and $\sigma^2(x)=\lambda(x)+\delta(x)$ is the diffusion coefficient. Equation (2.3) is the *Fokker-Plank equation* (FPE) for the analyzed process.

Diffusion model (2.3) could be considered as a limit of birth-death processes under some "scaling conditions". Roughly, let $\Delta x$ be a step size for a random process with the steps taken at time intervals $\Delta t$. Let $N$ be the number of steps and $t=N\Delta t$. If the steps are independent and uniformly distributed and both the mean displacement and the mean of square displacement, $\langle\Delta x\rangle$ and $\langle\Delta x^2\rangle$ are of the order of $\Delta t$, such that $\langle\Delta x\rangle/\Delta t \to \mu(x)$ and $\langle\Delta x^2\rangle/\Delta t \to \sigma^2(x)$ at $\Delta t \to 0$, then the probability density function (pdf) of the process can be approximated by the pdf of the diffusion process with the drift $\mu(x)$ and diffusion coefficient $\sigma^2(x)$. Conversely, for given functions $\mu(x)$, $\sigma^2(x)$, consider a discrete-parameter birth-and-death chain on $S=\{0,\pm\Delta,\pm 2\Delta,...\}$ with step size $\Delta$, time unit $\tau$ and transition probabilities $p_{i,i-1}(\Delta)= \sigma^2(i\Delta)\tau/(2\Delta^2) - \mu(i\Delta)\tau/(2\Delta)$, $p_{i,i+1}(\Delta) = \sigma^2(i\Delta)\tau/(2\Delta^2) + \mu(i\Delta)\tau/(2\Delta)$, $p_{i,i}(\Delta)=1-\sigma^2(i\Delta)\tau/\Delta^2$. Suppose that the sequence $\{\Delta_n, \tau_n\}$ with $\Delta_n \to 0$, $\tau_n \sim \Delta_n^2$ can be chosen in such a way that all quantities $p_{i,i-1}(\Delta_n)$, $p_{i,i+1}(\Delta_n)$, $p_{i,i}(\Delta_n)$ are non-negative. Then the corresponding sequence of the birth-and-death chains converges in distribution to a diffusion with drift $\mu(x)$ and diffusion coefficient $\sigma^2(x)$ (see, e.g., (17), ch.5.4 for details). Some variations of the birth-death type processes cannot be described by the simple diffusion equation, e.g., those that produce "anomalous diffusion" with a different scaling behavior. In general, the problem of "equivalence" of models (2.1) and (2.3) is non-trivial (see, e.g., (18,20,21)) and the models may have different dynamics.

In this work, our goal is to develop the appropriate mathematical approaches and tools for modeling the genome evolution rather than to investigate the equivalence of different mathematical approaches; our starting point is not a particular model but empirical data, such as the Pareto distribution of the sizes of the paralogous gene families. Therefore, we prefer to consider the FPE (2.3) as a distinct mathematical approach (although connected to the previous ones) to modeling genome evolution rather then as a technical approximation of the initial birth-and-death process with discrete set of states.

Let us denote $J(x,t)$ the *current* of "particles" through the point $x$ at time $t$

$$J(t,x) = f(t,x)\mu(x) - \frac{\partial}{\partial x}[\frac{1}{2} f(t,x)\sigma^2(x)]. \qquad (2.4)$$

Then the Fokker-Plank equation (2.3) can be written as an equation of continuity:

$$\frac{\partial f(t,x)}{\partial t} = -\frac{\partial}{\partial x} J(t,x). \qquad (2.5)$$

The FPE is a second-order parabolic differential equation, and to obtain the solution, we need an initial condition and boundary conditions at the ends of the interval $[r,N]$. For example, if a "particle" (in terms of our model, a gene family) cannot leave the interval (at least one member of the family is essential and the maximum possible size of the family is bounded) and there is zero net flow across the ends, then

$$J(t,x) = f(t,x)\mu(x) - \frac{\partial}{\partial x}[\frac{1}{2} f(t,x)\sigma^2(x)] = 0 \text{ at } x=r \text{ and } x=N. \qquad (2.6)$$

If the system is "open" at the left end $x=r$, and innovation is possible, then the rate of innovation or the current through the boundary point $J(t,r)$ can be taken as a boundary condition for (2.5).

## 3. Stationary solution of the model and the power asymptotics

The stationary solution $f_{st}(x)$ of the model (2.3) or (2.5), for which $df_{st}(x)/dt=0$, satisfies the equation $dJ(x)/dx=0$, so the current $J(x)$=const at all $x$. If the system is closed and hence $J(x) =0$ at $x=r$ due to the boundary condition, then $f_{st}(x)\mu(x) = \frac{1}{2}\frac{\partial}{\partial x}[f_{st}(x)\sigma^2(x)]$ for all $x\in[r,N]$. This equation can be easily solved (see, e.g., (18), ch.5):

$$f_{st}(x) = \frac{\sigma^2(r)f_{st}(r)}{\sigma^2(x)}\exp(2\int_r^x \frac{\mu(y)}{\sigma^2(y)}dy). \tag{3.1}$$

In the context of the present work, the case of interest is the power asymptotic of the stationary solution $f_{st}(x)$. Typically, $\sigma^2(x)$ is smaller or at most of the same order as $\mu(x)$; for example, if $\sigma^2(x)= \sigma^2$=const and $\mu(x)=cx$, $c<0$ is a constant, then the stationary distribution (3.1) is the (truncated) normal distribution; if $\sigma^2(x)=\sigma^2$=const and $\mu(x)=c<0$, then the stationary distribution (3.1) is the (truncated) exponential distribution. The power asymptotic of the stationary solution (3.1) appears only in the opposite case, i.e., when $\sigma^2(x)$ increases faster then $\mu(x)$. More precisely, let $\sigma^2(x)= x^\rho(a+o(1/x))$, $\rho$, $a>0$ are constants; let $\mu(x)/\sigma^2(x)= -\eta/(2x)+O(1/x^2)$; then $f_{st}(x)\sim x^{-\eta-\rho}$.

A more general assertion is given in (16), Theorem 1. As a representative example, let us consider a *linear* diffusion model with $\lambda(x)$ and $\delta(x)$ being linear functions of $x$: $\lambda(x)=\lambda(x+a)$, $\delta(x)= \delta(x+b)$ where $\lambda$ and $\delta$ are positive constants. The linear discrete-state BDIM (10) has a stable distribution $f_{st}(x)$ which is asymptotically equal to the power-law distribution if and only if $\lambda=\delta$; then $f(x)\sim x^{-\gamma}$, where $\gamma=b-a+1$. An analogous result is valid for the corresponding diffusion linear model with

$\mu(x) = (\lambda-\delta)x+\lambda a-\delta b$, $\sigma^2(x) =(\lambda+\delta)x+\lambda a+\delta b$:

$$\frac{\partial f(x,t)}{\partial t}=\{-\frac{\partial}{\partial x}[f(x,t)((\lambda-\delta)x+\lambda a - \delta b)]+\frac{1}{2}\frac{\partial^2}{\partial x^2}[f(x,t)((\lambda+\delta)(x+s)]\} \tag{3.2}$$

with $s=(a\lambda+b\delta)/(\lambda+\delta)$. For the linear diffusion model (3.2),

$f_{st}(x)=f_{st}(r)\exp(-l(x-r))(\frac{s+x}{s+r})^{-\gamma}$ where $\gamma=4(b-a)\frac{\lambda\delta}{(\lambda+\delta)^2}+1$, $l=2(\delta-\lambda)/(\lambda+\delta)$.

In particular, the stationary distribution of the linear diffusion model is the Pareto distribution if and only if $\lambda=\delta$ and $a-b\neq 1$; under these conditions,

$f_{st}(x)=c_1(s+x)^{-\gamma}$ where $\gamma=b-a+1$, $s=(a+b)/2$ and $c_1=f_{st}(r)(s+r)^{\gamma}$.

Next, let us consider a more general case and suppose that birth and death rates are polynomials on $x$. Such models can take into account interactions between particles and, in the case of gene families, reflect a feedback between the family size and growth rate; polynomial models with a discrete space of states were analyzed in (13,14). The asymptotic behavior of the stationary distribution critically depends on the relation between the degrees of the polynomials. In particular, it can be proved, based on the formula (3.1), that *the stationary distribution of the polynomial diffusion model asymptotically follows the power distribution if and only if the birth and death rates are polynomials of the same degree, n, with the same coefficients at $x^n$, $\lambda=\delta$. Then $f_{st}(x)\sim x^{-\gamma}$, where $\gamma=q_1-r_1+n$.*

## 4. Spatial-temporal dynamics of the linear model – a special class of solutions

The linear model (3.2) can be written in the form

$$\frac{\partial f(x,t)}{\partial \tau}=(1+\gamma)\frac{\partial f}{\partial x}+(x+s)\frac{\partial^2 f}{\partial x^2} \qquad (4.1)$$

where $\gamma=b-a+1$, $s=(a+b)/2$

This equation has a special family of solutions for any $t_0\geq 0$

$$f(x,t)=(x+s)^{-\gamma}[C_1+C_2\Gamma(\gamma,\frac{x+s}{\lambda(t+t_0)})], \qquad (4.2)$$

where $\Gamma(z,y)$ is the incomplete $\Gamma$–function. We will refer to the solutions of the form (4.2) as generalized self-similar (gss) solutions (see (16), Theorem 3).

Formula (4.2) describes the transformation of the initial stationary solution $f_{st}(x)=C_1(x+s)^{-\gamma}$ into another stationary solution of the same "Pareto shape",

$f_{st}(x)=(C_1+C_2\Gamma(\gamma))(x+s)^{-\gamma}$. Indeed, according to the known properties of incomplete $\Gamma$-function ((22), ch. 5), $\Gamma(\gamma,0)=\Gamma(\gamma)$, $\Gamma(\gamma,\infty)=0$. Hence, $f(x,0) = C_1(x+s)^{-\gamma}$, and $f(x,t)\to(C_1+C_2\Gamma(\gamma))(x+s)^{-\gamma}$ at $t\to\infty$.

The current for the gss-solution (4.2) is

$$J(t,x)= C_2 \exp\left(-\frac{s+x}{\lambda(t+t_0)}\right)(\lambda(t+t_0))^{-\gamma}. \tag{4.3}$$

If the current was zero at the initial instant, then $t_0=0$ (otherwise $t_0>0$) and

$$J(t,x)= C_2 \exp\left(-\frac{s+x}{\lambda t}\right)(\lambda t)^{-\gamma}.$$

The rate of increase of $f(t,x)$ at any point $x$ is

$$\frac{\partial}{\partial t}f(t,x)=-\frac{\partial}{\partial x}J(t,x)= J(t,x)/(\lambda(t+t_0)).$$

The rate of increase of $f(t,r_0)$ at the left boundary point $r_0$, which we call "influx rate", is

$$\frac{\partial}{\partial t}f(t,r_0)= -\frac{\partial}{\partial x}J(t,r_0)= J(t,r_0)/(\lambda t)= C_2 \exp\left(-\frac{s+r_0}{\lambda t}\right)(\lambda t)^{-\gamma-2}.$$

In particular, at $r_0=1$, the influx rate may be interpreted as the rate of increase of the number of singletons (families with a single member).

We can estimate the constants in solution (4.2) from the condition:

$$\int_1^N xf_{st}(x)dx = C\int_1^N x(x+s)^{-\gamma}dx=N_G \tag{4.4}$$

where $N_G$ is the total number of genes in recognizable families (the families were identified by detecting protein domains with the CDD collection of position-specific scoring matrices as previously described (10); it would be more precise to speak of domain families but we use the phrase gene families for the sake of simplicity), or from the condition

$$\int_1^N f_{st}(x)dx = C\int_1^N (x+s)^{-\gamma}dx=N_F \tag{4.4'}$$

where $N_F$ is the total number of families. In what follows we use (4.4) for the estimation of the constants using the previously reported $N_G$ and $N_F$ values; for example, $C=5436$ for *Dme* and $C=22160$ for *H. sapiens*. Hence, the constant $C_2$ in (4.2), for given $C$ and $C_1$, is $C_2=(C-C_1)/\Gamma(\gamma)$ for the linear model, for example, at $C_1=0$ we get $C_2=5014$ for *Dme* and

$C_2=19334$ for *H. sapiens*. The innovation rate, $v(t)$, at the boundary point $x=r_0$ can be defined from the relation

$\frac{\partial}{\partial t} f(t, r_0) = v(t) + \frac{\partial}{\partial x}[f(t,x)\delta(x) + 1/2 \frac{\partial}{\partial x} f(t,x) \sigma^2(x)]|x=r_0$, so

$v(t) = -\frac{\partial}{\partial x}[\lambda(x) f(t,x)]|x=r_0$.  (4.7)

At $r_0=1$, the innovation rate is equal to the rate of emergence of new families (singletons); this process provides the dynamics (increase or decrease) of the total number of families; in particular, it counterbalances the loss of singletons at the stationary state. For the linear model with $r_0=1$, the innovation rate in the stationary state $f_{st}(x) = C(x+s)^{-\gamma}$ is

$v = C\lambda (s+1-\gamma(1+a))(s+1)^{-\gamma-1}$.  (4.8)

The model parameter $\lambda$ can be estimated using the published estimates of the mean number of duplication, $L \sim 20/\text{Ga}$ years (15) as described previously for the discrete BDIMs (13). The stationary innovation rates for different species are given in Table 3.1. The innovation rates for the gss-solution of the linear diffusion model depending on time are shown in Fig. 1 for different species. In each case, the time dependence of the innovation rate has a distinct shape, with an initial accelerating phase and subsequent slow deceleration.

Formula (4.2) at $t_0=0$ describes the transformation of the initial stationary distribution $f_{st}(x) = C_1(x+s)^{-\gamma}$ into another stationary distribution $f_{st}(x) = (C_1+C_2\Gamma(\gamma))(x+s)^{-\gamma}$. Indeed, according to the properties of the incomplete $\Gamma$-function, $\Gamma(\gamma,0) = \Gamma(\gamma)$, $\Gamma(\gamma,\infty)=0$. Hence, in the initial instant $t=0$ (and, also, $t_0=0$) $f(x,t)=C_1(x+s)^{-\gamma}$, and $f(x,t) \to (C_1+ C_2\Gamma(\gamma))(x+s)^{-\gamma}$ at $t \to \infty$. The transition process is considered in greater detail below.

## 5. Transformation of the model and the FPE for the transformed model

The starting point of this analysis was the stationary distribution of gene family size which was extracted from the empirical data on prevalence of domains in proteins and followed the (truncated) Pareto distribution. Here we construct a class of models with a common, fixed stationary distribution but with different dynamics. A similar approach was realized previously for the stochastic discrete-space BDIMs with the birth rate $\lambda(x)$ and death rate $\delta(x)$ (13).

Let $g(x)$ be a positive smooth function; transform the initial birth and death rates $\lambda(x)$ and $\delta(x)$ using formulas

$\lambda^*(x) = \lambda(x)g(x) + (\delta(x) + \lambda(x))/4 \, \partial g(x)/\partial x = \lambda(x)g(x) + \sigma^2(x)/4 \, \partial g(x)/\partial x$,

$\delta^*(x) = \delta(x)g(x) - (\delta(x) + \lambda(x))/4 \, \partial g(x)/\partial x = \delta(x)g(x) - \sigma^2(x)/4 \, \partial g(x)/\partial x$.

The diffusion model can be considered an analog of a birth-and-death process only if $\lambda^*(x)$ and $\delta^*(x)$ are non-negative for $r < x < N$; accordingly, we suppose that the function $g(x)$ is such that $\lambda^*(x)$ and $\delta^*(x)$ are non-negative for $r < x < N$. Then it follows from formula (3.1) that

*the stationary solutions of the initial diffusion model* (2.3) *and transformed model with* $\sigma^{*2}(x) = \lambda^*(x) + \delta^*(x) = \sigma^2(x) g(x)$, $\mu^*(x) = \mu(x)g(x) + \sigma^2(x)/2 \, \partial g(x)/\partial x$ *are identical up to the normalizing constant.*

The stationary solution of the linear model is $f_{st}(x) = f_{st}(r)(x+s)^{-\gamma}$, $\gamma = b - a + 1$, so all formulas for the diffusion model, transformed by the function of the same power form, $g(x) = (x+s)^{\rho-1}$, are especially simple:

$$\lambda^*(x) = [(x+a)(x+s)^{\rho-1} + (x+s)(x+s)^{\rho-2}(\rho-1)/2] = (x+a+(\rho-1)/2)(x+s)^{\rho-1}, \quad (5.1)$$

$\delta^*(x) = [(x+b)(x+s)^{\rho-1} - (x+s)(x+s)^{\rho-2}(\rho-1)/2] = (x+b-(\rho-1)/2)(x+s)^{\rho-1}$,

$$\mu^*(x) = (\rho-\gamma)(x+s)^{\rho-1}, \quad \sigma^{*2}(x) = 2(x+s)^{\rho}. \quad (5.2)$$

In what follows we explore mainly the diffusion models with the drift and diffusion coefficients (5.2); these models are transformations of the linear diffusion model by the power function $g(x) = (x+s)^{\rho-1}$ and can be considered formal diffusion approximations of the BDIMs with birth and death rates (5.1). For the transformed linear diffusion model, the current is $J^*[f](t,x) = -\lambda g(x)[\gamma f(t,x) + (x+s)\frac{\partial}{\partial x} f(t,x)]$. Thus, the FPE for the transformed model is $\frac{\partial}{\partial t} f(t,x) = -\frac{\partial}{\partial x} J^*[f](t,x)$, or

$$\frac{\partial}{\lambda \partial t} f(t,x) = \frac{\partial}{\partial x} g(x)[\gamma f(t,x) + (x+s)\frac{\partial}{\partial x} f(t,x)].$$

If $g(x)=(x+s)^{\rho-1}$ (with $\rho \geq 1$, $b>(\rho-3)/2$), then the corresponding FPE is of the form

$$\frac{\partial}{\lambda \partial t} f(t,z) = z^\rho \left\{ \frac{\partial^2}{\partial z^2} f(t,z) + (\gamma+\rho)/z \frac{\partial}{\partial z} f(t,z) + [\gamma(\rho-1)/z^2] f(t,z) \right\} \qquad (5.3)$$

where $z=x+s$.

## 6. The mean first passage time: the mean time of formation of the largest family

As mentioned above, the discrete-space birth-and-death process and the corresponding diffusion model cannot be completely equivalent and hence, in general, might predict different regimes of genome evolution. Let us compare an important characteristic of genome evolution, the mean time required for a family to reach the largest possible size, for both models. According to the well-known results ((18), Ch.5.2, formulas (6.2.158), (6.2.160)), the mean first passage time for the diffusion process with a reflecting boundary $r$ and absorbing boundary $N$, $r<N$, starting from the state $x$, is

$$T(x) = 2 \int_x^N \frac{dy}{\psi(y)} \int_r^y \frac{\psi(z)}{\sigma^2(z)} dz, \text{ where } \psi(x) = \exp(2 \int_r^x \frac{\mu(y)}{\sigma^2(y)} dy).$$

For the diffusion model (5.3), $\psi(x) = (\frac{x+s}{r+s})^{\rho-\gamma-1}$. So for $\rho \neq 2$

$$T(x;\rho) = \frac{1}{\gamma(\gamma+2-\rho)(r+s)^\gamma} [(N+s)^{\gamma+2-\rho} - (x+s)^{\gamma+2-\rho}] - \frac{1}{\gamma(2-\rho)} [(N+s)^{2-\rho} - (x+s)^{2-\rho}]$$

and for $\rho=2$

$$T(x;2) = \frac{1}{\gamma^2(r+s)^\gamma} [(N+s)^\gamma - (x+s)^\gamma] + \frac{1}{\gamma} \ln[(x+s)/(N+s)].$$

Let us compare the mean largest family formation times for the original (discrete-space) non-linear BDIM and its diffusion approximation. Figure 2 shows that the formation time for the discrete-space model and that for the corresponding diffusion model are identical for $d<3$.

## 7. Temporal dynamics and gss-solutions for the transformed model

A solution $f(x,t)$ of FPE (2.3) is considered *generalized self-similar* (gss) if it is of the form $f(x,t) = x^a G(y)$ where $y = x/\phi(t)$ with smooth $\phi(t) \neq 0$ and $a$ is a (real) constant. In the context of modeling genome size evolution, we were interested mainly in bounded solutions of the model. The following crucial assertion can be proved:

*equation (5.3) for $\rho < 2$ has a family of gss-solutions*

$$f(t,x) = (x+s)^{-\gamma} \left[ C_1 + C_2 \Gamma\left(1 + \frac{\gamma-1}{2-\rho}, \frac{1}{(2-\rho)^2} \frac{(x+s)^{2-\rho}}{\lambda(t+t_0)}\right)\right] \quad (7.1)$$

where $C_1$, $C_2$ are arbitrary constants and $t_0 \geq 0$. These solutions are bounded functions of $t$ at $t \to 0$, $t \to \infty$ for any constants $C_1$, $C_2$.

*For $\rho = 2$, equation (5.3) has a family of gss-solutions*

$$f(x,t) = (x+s)^{-\gamma} \{C_1 + C_2 [(x+s)\exp(-\alpha\lambda(t+t_0))]^{\gamma-\alpha-1}\} \quad (7.2)$$

*which are bounded functions of $t$ at $t \to 0$, $t \to \infty$ for any value $0 < \alpha < \gamma - 1$ and any constants $C_1$, $C_2$.*

A detailed description of the gss-solutions of equation (5.3) for different $\rho$ values is given in (16) (see Theorems 3 and 4). For $\rho < 2$, due to known properties of the incomplete $\Gamma$-function, $\Gamma(1+\gamma/(2-\rho), 1/(2-\rho)^2 (x+s)^{2-\rho}/(\lambda t)) \to 0$ at $t \to 0$ and $\Gamma(1+\gamma/(2-\rho), 1/(2-\rho)^2 (x+s)^{2-\rho}/(\lambda t)) \to \Gamma(1+\gamma/(2-\rho))$ at $t \to \infty$. Thus, formula (7.1) at $t_0 = 0$ describes the transition from the initial stationary distribution $f(0,x) = C_1(x+s)^{-\gamma-1}$ to the final stationary distribution $f(\infty,x) = (x+s)^{-\gamma}[C_1 + C_2\Gamma(1+(\gamma-1)/(2-\rho))]$ which differs from the initial one only by a constant multiplier.

In the case of $\rho = 2$, the behavior of gss-solutions can substantially change. Any solution (7.2) is bounded only if $0 \leq \alpha < \gamma + 1$. In this case, the solution describes the transition from the initial stationary distribution $f(0,x) = C_1(x+s)^{-\gamma} + C_2(x+s)^{-\alpha-1}$ to the final stationary distribution $f(\infty,x) = C_1(x+s)^{-\gamma}$. Let us note that the initial and final solutions now have different shapes.

For $2 < \rho < 1 + \gamma$, the *gss-solution is unbounded at $t \to \infty$*; typically, according to our results with discrete-space BDIM (10), $\gamma > 2$. We further explore some important peculiarities of the gss-solutions. As in the case of the linear diffusion model, formula

(7.1) describes the process of transformation of the initial stationary solution $f_{st}(x)=C_1(x+s)^{-\gamma}$ into another stationary solution of the same "Pareto shape",

$$f_{st}(x)=(x+s)^{-\gamma}[C_1+ C_2\Gamma(1+\frac{\gamma-1}{2-\rho})].$$

For the numerical estimations, we choose the values of constants $C_1$, $C_2$ such that the total number of genes ($N_G$) in the stationary state of the model, computed using formula (4.4), was equal to the observed $N_G$ value. For example, for *H. sapiens* $N_G=27844$ and $C=22160$ and $C_2= (C-C_1)/\Gamma(1+\frac{\gamma-1}{2-\rho})$. Additionally, the value of the model parameter $\lambda$ has to be computed such that agreement with available estimates of the gene duplication rate (15) is reached (see (13) for details). The mean rate of gene duplication in the stationary state, according to the diffusion model and formula (5.1) for the birth rate $\lambda^*(x)$, is equal to

$$\lambda\int_1^N \frac{x+a+(\rho-1)/2}{x}(x+s)^{-\gamma-1+\rho}dx / \int_1^N (x+s)^{-\gamma}dx = 20 \text{ per 1 Ga},$$ where $N$ is the maximum size of gene families. Hence, the value of the parameter $\lambda$ depends on the model degree $\rho$. Figure 3 shows two stationary solutions (which differ only by a multiplying constant) and the process of evolution of one into the other, i.e., the solutions (7.1 for $t_0=0$) at different time moments. If the constant $C_1=0$ in formula (7.1), then this solution describes the emergence of the stationary distribution triggered by the innovation process, i.e., the current through the left boundary (Figure 4).

The rate of increase of $f(t,r_0)$ at the left boundary point $r_0=1$, which we called influx rate and interpreted as the rate of increase of the number of singletons is shown on Fig. 5. The process of innovation or, more precisely, the innovation rate taken as the boundary condition specifies the solution of the model; in particular, the gss-solution (7.1) corresponds to the innovation rate $v(t)=-\frac{\partial}{\partial x}[\lambda(1)f(t,1)]$ with $\lambda(x)=(x+a+(\rho-1)/2)(x+s)^{\rho-1}$ according to formula (5.1) and $f(t,x)$ defined by (7.1). The innovation rates for the gss-solutions of the diffusion model of different degrees depending on time are shown in Figures 6 and 7.

It should be emphasized that the gss-solutions describe the regime of genome size increase with time; different points of the curve move with different speeds such that the shape of the curve changes with time and the curves at different instants are not exactly self-similar. It is interesting to estimate the speed of the movement. To do so, let us fix a point $(x_1, f_1)$ on the curve $f(t_1,x)=y$ at the instant $t_1$, so that $f(t_1,x_1)= f_1$ and trace the movement of the point $x(t)$ defined by the equation $f(t,x(t))= f_1$. Figure 8 shows the values $x(t)$: $f(t,x(t))=0.1$ for the diffusion models of different degrees.

The speed of genome growth decreases when the genome size approaches its stationary value. Importantly, for $\rho>1$, there exist initial stages when the genome size increases with acceleration. Figure 9 shows the speed of movement of the point $x(t)$: $f(t,x(t))=0.1$ for models of different degrees. We can see that the speed of the genome growth increases during approximately the first 1 Ga years.

The gss-solution (7.1) describes the process of formation of the stationary Pareto-like solution at $t\to\infty$. To clarify the structure of the transition regimes at time moments $t<\infty$, let us consider the asymptotic of the gss-solution at large $x$ under fixed $t$. It can be shown that

$$f(t, x) \approx C_1 (x+s)^{-\gamma} + \qquad (7.3)$$
$$C_2(\rho)(\lambda t)^{(\gamma-1)/(2-\rho)} (x+s)^{-1} \exp[-(x+s)^{2-\rho}/((2-\rho)^2\lambda t)] [1+(\gamma-1)(2-\rho)(\lambda t)/(x+s)^{2-\rho}]$$

for large $x$ with a good accuracy if $(\gamma-1)/(2-\rho)<3$, or $\rho<2-(\gamma-1)/3$. For example, according to our results, $\gamma=2.27$ for *H. sapiens*, hence the expansion (7.3) is valid at $\rho<1.57$. Thus, if $C_1=0$ (no genes at the initial moment), at any fixed moment of time, the gss-solution as a function of $x$ is of the form

$$f(t, x) \approx c_1(\rho,t) (x+s)^{-1} \exp(-c_2(\rho,t) (x+s)^{2-\rho}) [1+ c_3(\rho,t)/(x+s)^{2-\rho}].$$

In particular, for the linear model with $\rho=1$,

$$f(t, x) \approx c_1(t) (x+s)^{-1} \exp(-c_2(t) (x+s)) [1+ c_3(\rho,t)/(x+s)].$$

Let us recall that, for the linear diffusion BDIM, $f_{st}(x)=f_{st}(r) \exp(-l(x- r)) (\frac{s+x}{s+r})^{-\gamma}$.

We can see that, at any fixed instant $t,$ the gss-solution

$$f(t,x) \approx c_1(t)(x+s)^{-1} \exp(-c_2(t) (x+s))$$

approximately coincides with the stationary solution of the linear (and even simple) *non-balanced* BDIM with

$\lambda(x) = \lambda(x+a)$, $\delta(x) = \delta(x+a)$ for any constant $a$ and $\lambda$, $\delta$ such that $2(\delta-\lambda)/(\lambda+\delta) = c_2(t)$.

## 8. Conclusions

We constructed and explored a class of continuous-state models whose dynamics is described by the Fokker-Plank equation and the stationary solution can be any specified Pareto function. These models are the diffusion version of the discrete-space BDIM that have been analyzed and applied to genome evolution in our previous studies. These diffusion models have *time-dependent* solutions of a special kind, namely, the gss-solutions (in general, such solutions are not known for the discrete-space BDIM); for the diffusion approximation, we found these solutions in the exact and explicit forms. The gss-solutions appear only with specific boundary conditions which, in the context of genome evolution, can be expressed in terms of the innovation rate. Preliminary computer simulations (GPK and FSB, unpublished) show that the gss-solutions are stable, i.e., small deviations from the boundary conditions lead to solutions that tended to the respective gss-solution at $t \to \infty$.

It should be emphasized that, for gss-solutions, the current $J(t,x) \neq 0$ at $t < \infty$ for any $0 < x < \infty$ (see formula (4.3)); therefore, gss-solutions exist only for the unbounded phase space $[r_0, \infty)$. The number of paralogous families in any genome is, obviously, finite; thus, the gss-solutions should be viewed as a mathematical approximation of genome evolution which, however, gives highly accurate results as long as the distribution of family sizes is described by a Pareto function.

The discrete-space BDIMs allowed us to identify the model parameters that provide for the evolution of the observed distributions of paralogous family sizes within a realistic timeframe. However, analytical, time-dependent solutions are not known for this class of models. In contrast, such solutions were obtained for the diffusion models. Interestingly, the gss-solutions of these models reveal a biphasic curve of genome growth, with the initial, relatively short, self-accelerating phase followed by a prolonged phase of slow deceleration. This regime of evolution was observed when genome growth started from zero (no genes at all) and proceeded via innovation, which may be interpreted as modeling primordial evolution, and also when evolution was allowed to proceed from one stationary state to another. The diffusion BDIM are highly abstract models and biological implications should be considered with utmost caution. Nonetheless, it is tempting to interpret the evolutionary regime derived from the gss-solutions as a punctuated-equilibrium-like phenomenon (23,24) whereby evolutionary transitions are accompanied by rapid gene amplification and innovation, followed by slow relaxation to a new stationary state.

Figure legends

Figure 1. Innovation rates for the gss- solution of the linear diffusion model. Species: *D. melanogaster* (blue), *H. sapiens* (red), *A. thaliana* (green), *C. elegans* (purple).

Figure 2. The formation time of the largest family for the discrete-space models (red) and for the corresponding diffusion models (black) depending on the model degree *d* (the model coefficients are taken for *H. sapiens*).

Figure 3. Evolution of the initial stationary solution (green) into the final one (red); the gss-solutions (7.1) of the diffusion model with $\rho=1.5$, $C_1=50$ and $C_2=6364$ is shown at $t=$ 0.1, 0.5, 1, 2, 5, 10 (blue; from down-left to up-right; double logarithmic scale).

Figure 4. Formation of the stationary solution (red) via innovation; the gss-solutions (7.1) of the diffusion model with $\rho=1.5$, $C_1=0$ and $C_2=6378$ are shown at $t=$ 0.1, 0.2, 0.5, 1, 2, 5,10 (blue; from down-left to up-right; double logarithmic scale).

Figure 5. Influx rate for the diffusion models with $\rho=1$ (black), $\rho=1.3$ (green), $\rho=1.5$ (blue), $\rho=1.6$ (red). The model parameters are taken for *H. sapiens*.

Figure 6. Innovation rate for the non-linear diffusion models with $\rho=1, 1.2, 1.5, 1.7$ (from top to bottom). The model parameters are taken for *H. sapiens*.

Figure 7. Dynamics of innovation rate for the non-linear diffusion models depending on the model degree. The model parameters are taken for *H. sapiens*.

Figure 8. The movement of the gss-solutions (at the point $f=0.1$) for models of different degrees: d=1 (black), d=1.5 (green), d=1.65 (blue), d=1.75 (red).

Figure 9. Speed of movement of the gss-solution (at the point $f=0.1$) for models of different degrees: d=1 (black), d=1.5 (green), d=1.65 (blue), d=1.75 (red).

**Table 1. Innovation rates for the gss-solutions of the linear diffusion model**

| Species | $N$ | $N_F$ | $N_G$ | $C$ | Maximum innovation rate (events/Ga) | $t_{max}$ | Stationary innovation rate | $\lambda$ |
|---|---|---|---|---|---|---|---|---|
| Dme | 335 | 1405 | 11734 | 5436.0 | 5179.0 | 0.21 | 4571.0 | 12.82 |
| Cel | 662 | 1418 | 17054 | 2206.0 | 3872.2 | 0.145 | 3020.0 | 14.65 |
| Ath | 1535 | 1405 | 21238 | 10048.3 | 2615.0 | 0.487 | 2369.0 | 9.85 |
| Hsa | 1151 | 1681 | 27844 | 22160.0 | 2838.4 | 0.71 | 2629.4 | 8.69 |

$N$, maximum size of a family; $N_F$, number of families; $N_G$, number of genes; $C$, normalized constant for the stationary state; $t_{max}$, time at which the maximum innovation rate is reached; *Stationary innovation rate*, the innovation rate in the stationary state; $\lambda$, the model parameter.

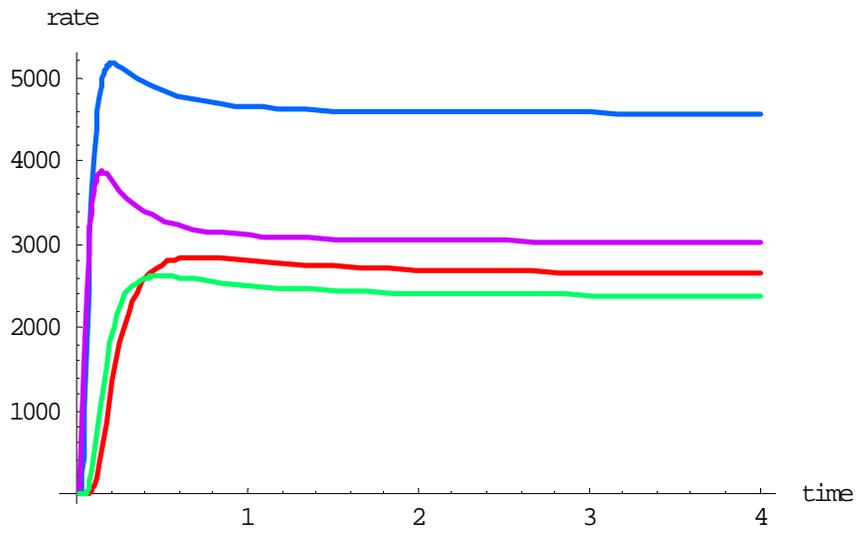

Fig.1

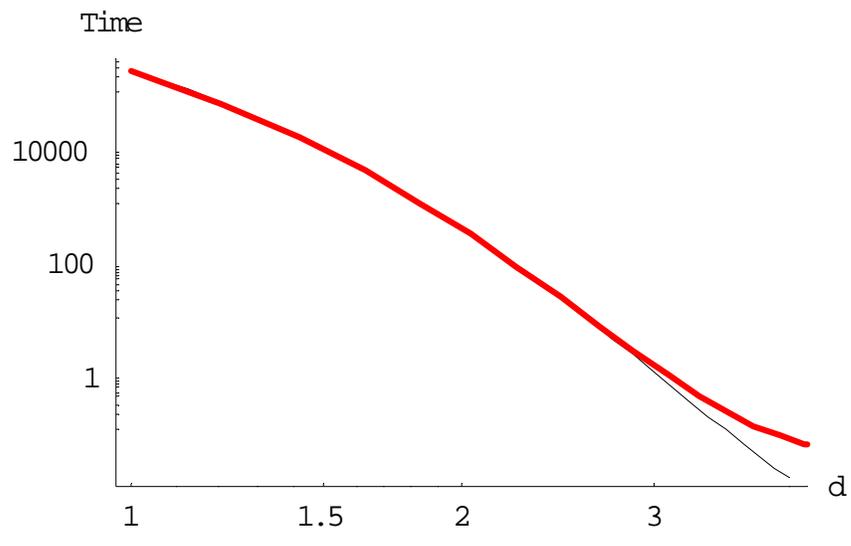

Fig.2

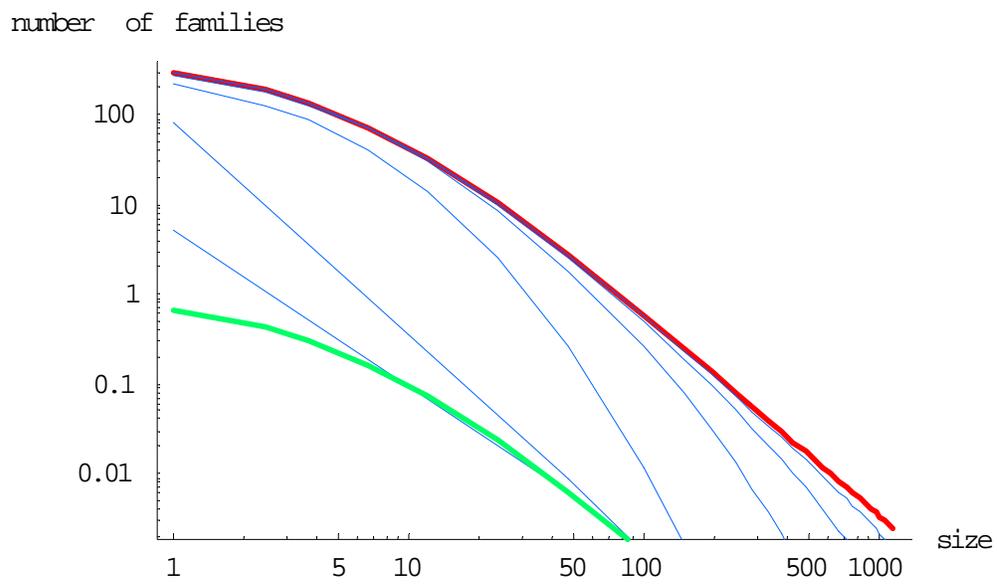

Fig.3

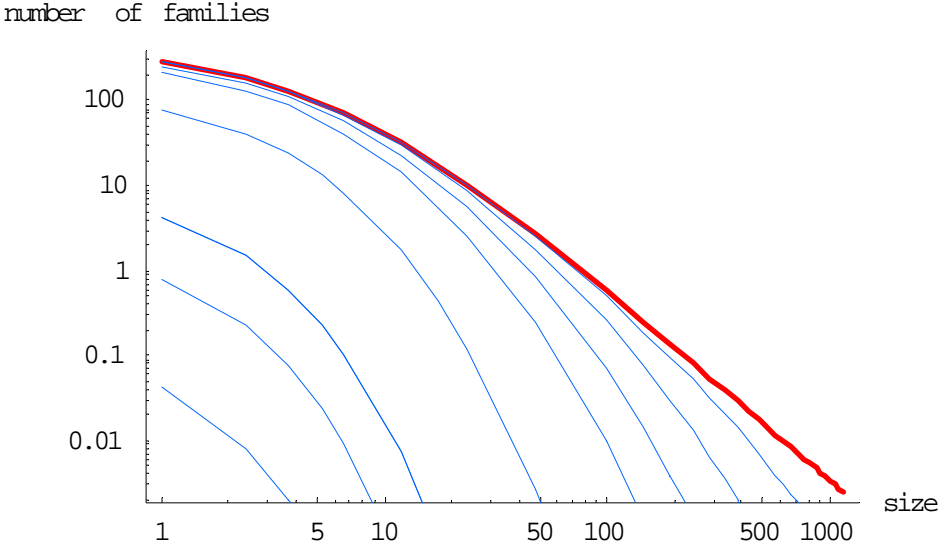

Fig.4

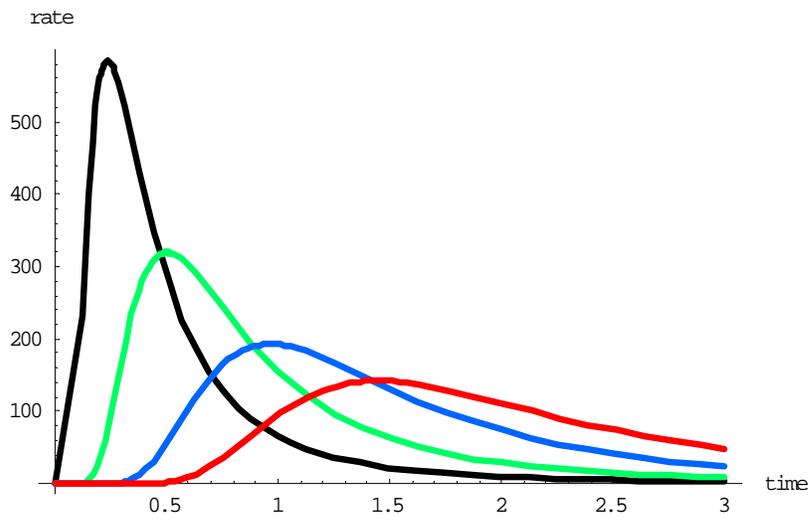

Fig. 5

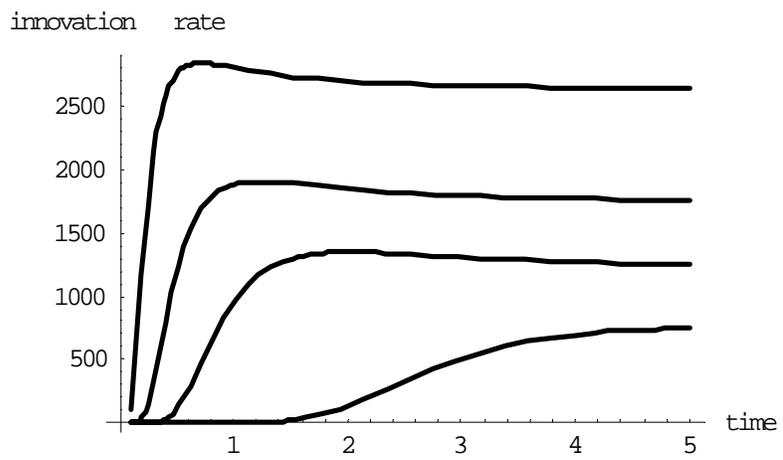

Fig.6

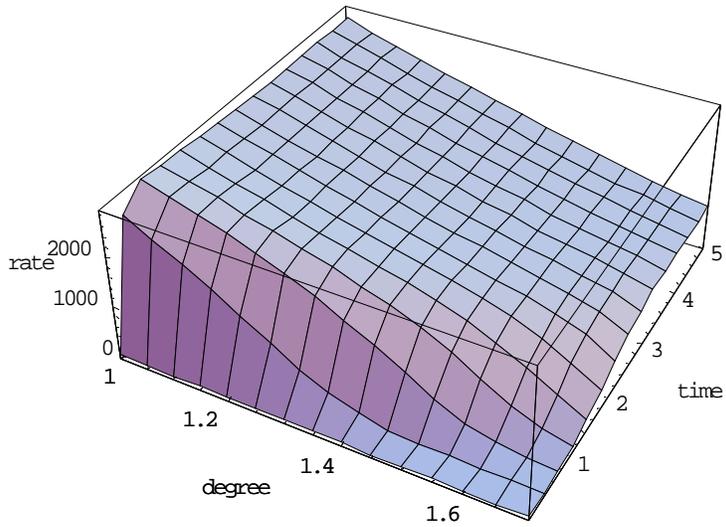

Fig. 7

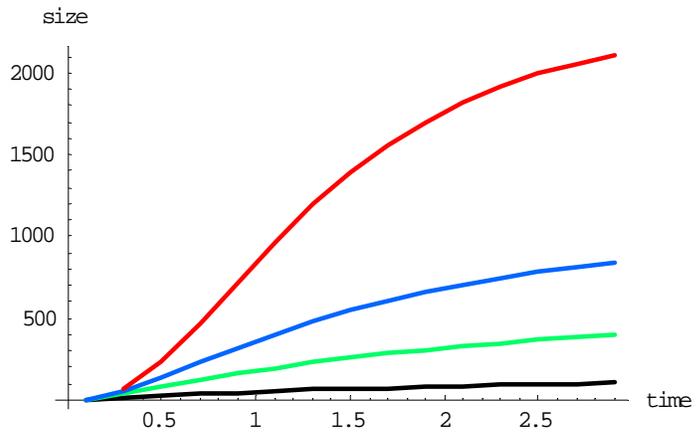

Fig.8

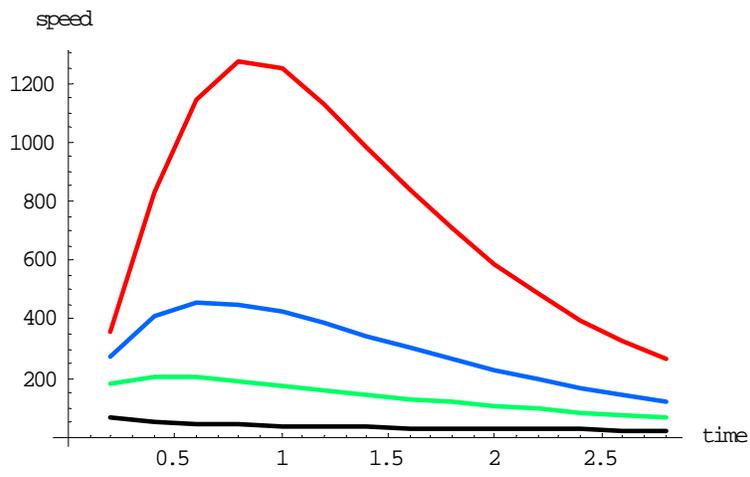

Fig.9